\documentclass[conference]{IEEEtran}
\usepackage{epsfig}
\usepackage{graphicx}
\usepackage{amsmath}
\usepackage{amssymb}
\usepackage{verbatim}
\usepackage{mathrsfs}
\usepackage{algorithm}
\usepackage[noend]{algpseudocode}
\usepackage{xcolor}
\usepackage{colortbl}
\usepackage{lipsum}
\usepackage{amsmath,amssymb}
\usepackage{amssymb}%
\usepackage{mathrsfs}%
\usepackage{blindtext}

\makeatletter
\def\BState{\State\hskip-\ALG@thistlm}
\makeatother
\usepackage{cite}
\usepackage[T1]{fontenc}

\IEEEoverridecommandlockouts
\begin{document}
%
\title{Deep Reinforcement Learning for Resource Allocation in V2V Communications}
\author{\IEEEauthorblockN{Hao Ye and Geoffrey Ye Li
}\\

\IEEEauthorblockA{School of Electrical and Computer Engineering\\
Georgia Institute of Technology\\
Email: yehao@gatech.edu; liye@ece.gatech.edu}

\thanks{The work of H. Ye and G. Y. Li was supported in part by a research
gift from Intel Corporation and the National Science Foundation under Grants
1405116 and 1443894.}
}
\maketitle
%
\begin{abstract}
In this article, we develop a decentralized resource allocation mechanism for vehicle-to-vehicle (V2V) communication systems based on deep reinforcement learning. Each V2V link is considered as an agent, making its own decisions to find optimal sub-band and power level for transmission. Since the proposed method is decentralized, the global information is not required for each agent to make its decisions, hence the transmission overhead is small. From the simulation results, each agent can learn how to satisfy the V2V constraints while minimizing the interference to vehicle-to-infrastructure (V2I) communications.
\end{abstract}
\section{Introduction}
\label{sec:intro}

Vehicle-to-vehicle (V2V) communications have become an important technology for improving transportation services and reducing road casualties.
Due to vital applications in the traffic safety, the requirements for the V2V communication links are often very stringent, i.e., the millisecond of end-to-end latency and nearly 100\% of reliability \cite{V2X_service}, which has raised a lot of attention both in academic and industry.
The Third Generation Partnership (3GPP) supports V2V services based on device-to-device (D2D) communications\cite{3GPP} since D2D shows superior performance to satisfy the quality-of-service (QoS) requirement of V2V applications.

In D2D systems, effective resource management needs to properly coordinate mutual interference between the cellular and the D2D users.
A three-step approach has been proposed in \cite{Feng} to control transmission power and allocate spectrum to maximize system throughput with
constraints on minimum signal-to-interference-plus-noise ratio (SINR) for both the cellular and the D2D links.
In V2V systems, new challenges have been brought by the high mobility vehicles, which causes wireless channels to change rapidly over time.
Therefore, the traditional methods on resource management for D2D communications with full channel state information (CSI) assumption can no longer be applied since it would be hard to track channel variations on such a short timescale.

There have been many interesting works on resource allocation for D2D based V2V communications.
Most of them are centralized,  where the central controller collects information and makes decisions for the vehicles.
With the global information of the networks, resource allocation can be formulated as optimization problems, where the QoS requirements of V2V serve as constraints.
However, these problems are usually NP-hard and therefore difficult to solve even with the global information of the networks.
As a result, various simplified approaches have been proposed to decompose the problems into multiple steps so that local optimal or sub-optimal solutions can be found.
In \cite{Sun}, the reliability and latency requirements of vehicular communications have been converted into optimization constraints, which are computable with only large-scale fading information and a heuristic approach has been developed to solve the resource management optimization problem.
In \cite{Le}, a resource allocation scheme has been designed based on slowly varying large-scale fading information only, where the sum V2I ergodic capacity is maximized with V2V reliability guaranteed.

Nevertheless, centralized control schemes will incur a large transmission overhead to get the global network information, growing linearly with mobile speed and quadratically with the number of vehicles. Thus they are not applicable to large networks.
Recently, some decentralized resource allocation mechanisms for V2V communication systems have been developed.
In \cite{baseline}, a distributed approach has been proposed to allocate sub-band to the V2V link by the position information. The V2V links are first clustered based on the positions of the vehicles and load similarities.
The resource blocks (RBs) are assigned to each group, and then in each group the assignments are refined through iterative swap within each group rather than in the whole network.
The low-complexity algorithm in \cite{Low_Com} optimizes outage probabilities for V2V communications based on bipartite matching.

In the previous works, the latency constraint for V2V links has not been considered too much since it is hard to be modeled directly into the optimization problems.
In order to address the problems that the traditional methods lack the ability to handle, in this article, we apply the multi-agent deep reinforcement learning scheme for resource allocation in V2V communication systems.
Reinforcement learning solves problems where each V2V link, as an agent, learns to make optimal decisions on spectrum and power for transmission based on the interacting with the environment. By optimizing strategies from the experience, the reward, which is a function of the capacity of the V2V link and the corresponding latency, is maximized in the long run.

Recently, deep learning has made great success in computer vision \cite{Imagenet}, speech recognition \cite{DL_Speech}, and wireless communications \cite{Hao}. With the help of deep learning techniques, reinforcement learning has shown impressive improvement in applications, such as playing videos games\cite{Atari} and Go games\cite{Go}.
Deep reinforcement learning has also been applied in resource allocation.
A deep reinforcement learning based approach has been proposed in \cite{DRL_RM} to address the problem of job scheduling with multiple resource demands, where the objective is to minimize the average job slowdown and the reward function is based on the reciprocal duration of the job.

In our system, deep reinforcement learning is used to find the mapping between the partial observations of each vehicle and the resource allocation solution. Each V2V link is considered as an agent and the channels and transmission power are selected based on the observations of instantaneous channel conditions and exchanged information shared from the neighbors at each time slot. In general, the agents will automatically balance between minimizing the interference of V2V links to the V2I networks and transmission power of V2V links to meet the requirements for the V2V link constraints, such as the latency constraints.

The main contribution of this article is using multi-agent deep reinforcement learning to develop a decentralized resource allocation mechanism for V2V communications, where the constraints on latency can be directly addressed.
Based on our the simulation results, deep reinforcement learning can learn to share the channel with other V2V links and generate the least interference to the V2I channels.

\section{System Model}

\begin{figure}[!t]
\centering
\includegraphics[width=0.9\linewidth]{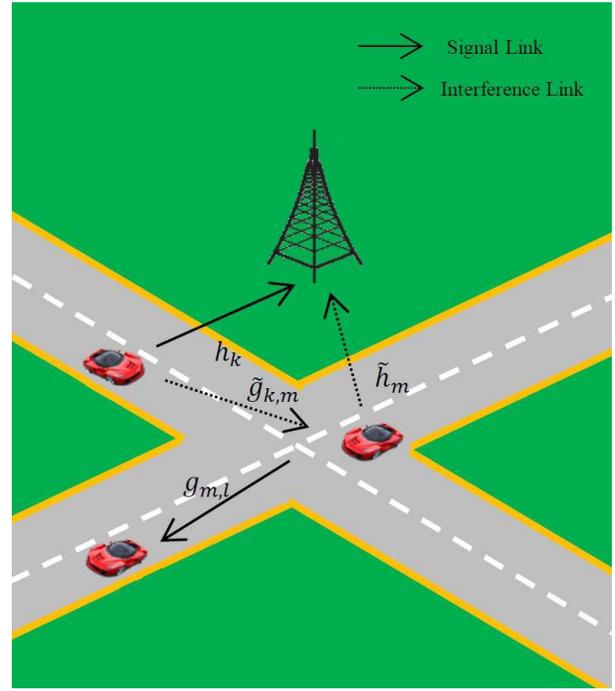}
\caption{V2V and V2I links }
\end{figure}

In this section, we will introduce the model of vehicle communications networks.
As shown in Fig. 1, the V2X networks consists of $\mathcal{M} = \{1,2,...,M\}$ cellular users (CUEs) demanding V2I links, which are orthogonally allocated spectrum and with high capacity communication links, and $\mathcal{K} = \{1,2,...,K\}$ pairs of D2D users (DUEs), which need V2V links to share information for traffic safety.
In order to improve the spectrum utilization efficiency, orthogonally allocated uplink spectrum for V2I link is reused by the V2V links since uplink resources are less intensively used and interference at the base station (BS) is more controllable.

The interference to the V2I links consists of two parts: the background noise and the signal from the V2V links sharing the same sub-band.
The SINR for the $m$th CUE will be
 \begin{equation}
\gamma_m = \frac{P_c h_{m}}{\sigma^2 + P_d \sum_{k\in \mathcal{K}}{\rho_{m,k} \tilde{h}_{k}}},
\end{equation}
where $P_c$ and $P_d$ are the transmission powers of CUE and DUE, respectively, $\sigma^2$ is the noise power, $h_{m}$ is the power gain of the channel corresponding to the $m$th CUE, $\tilde{h}_{k}$ is the interference power gain of the $k$th DUE, and $\rho_{m,k}$ is the spectrum allocation indicator with $\rho_{m,k} = 1$ if the $k$th DUE reuses the spectrum of the $m$th CUE and $\rho_{m,k} = 0$ otherwise.
Hence the capacity of the $m$th CUE can be expressed as
\begin{equation}
 C_m = W \cdot \log(1+\gamma_m),
\end{equation}
 where $W$ is the bandwidth.


Due to the essential role of V2V communications in vehicle security protection, there are stringent latency and reliability requirements for V2V links while the data rate is not of great importance.
Traditionally, these constraints are handled by converting into the outage probabilities \cite{Le,Sun}.
In our method, the latency and reliability constraints are modeled in the reward function directly, where a lower reward is given when the constraints are violated.

In contrast to V2V safety communications, the latency requirement is less strict for the traditional cellular traffic.
Therefore, traditional resource allocation focuses on maximizing the throughput under certain fairness considerations.
The maximization of the V2I sum rate will be reflected in the reward function in our method.

Since we are developing a decentralized control approach for the network, the BS is assumed to have no information on the V2V links.
As a result, the resource allocation procedures of the V2I network should be independent of the V2V links.
Given the V2I links, the main goal of the proposed autonomous scheme for joint channel and power level allocation is to ensure the latency constraints for each V2V link.

\section{Deep Reinforcement Learning for Resource Allocation}
In this section, the framework on deep reinforcement learning for resource allocation in V2V communications is introduced, including how to represent the key parts in the reinforcement learning framework and how to train the deep Q-networks.

\subsection{Reinforcement Learning }
\begin{figure}[!t]
\centering
\includegraphics[width = 0.9 \linewidth]{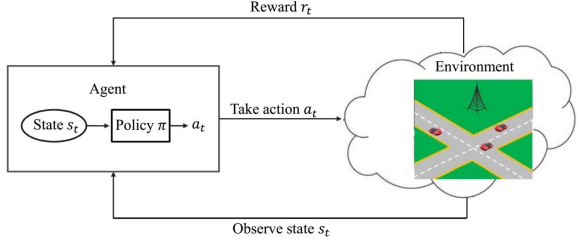}
\caption{Deep reinforcement learning for V2V communications}
\end{figure}

The structure of reinforcement learning for V2V communications is shown in Fig. 2, where an agent, corresponding to a V2V link, interacts with the environment.
In this scenario, the environment is considered to be everything outside the V2V link.
It should be noted that the behaviors of other V2V links cannot be controlled in the decentralized settings. As a result, their actions, such as selected spectrum, transmission power, etc., are treated as a part of the environment.

At each time $t$, the V2V link, as the agent, observes a state, $s_t$, from the state space, $\mathcal{S}$, and accordingly takes an action, $a_t$, from the action space, $\mathcal{A}$,  selecting sub-band and transmission power based on the policy, $\pi$. The decision policy, $\pi$, is determined by a Q-function, $Q(s_t, a_t, \theta)$, where $\theta$ is the parameter of the Q-function and can be obtained by deep learning.
Following the action, the state of the environment transitions to a new state $s_{t+1}$ and the agent receives a reward, $r_t$, determined by the capacity of the V2V link and the corresponding latency.
In our system, the state observed by each V2V link for characterizing the environment consists of several parts: the instant channel information of the V2V corresponding link, $g_t$, the previous interference to the link, $I_{t-1}$, the channel information of the V2I link, e.g., from the V2V transmitter to the BS, $h_t$, the selected of sub-channel of neighbors in the previous time slot, $S_{t-1}$, the remaining load of the DUE to transmit, $L_t$ , and the remaining time to meet the latency constraints $R_t$. Hence the state can be expressed as $s_t = [g_t, I_{t-1}, h_t, S_{t-1}, L_t, R_t]$. The instant channel information and the interference received reveal the quality of each sub-band. The distribution of neighbors' selection relates the interference to the other DUE users. The remaining amount of message to transmit and the remaining time contain information for selecting suitable power level.

At each time, the agent takes an action $a_t \in \mathcal{A}$, which includes selecting a sub-channel and a power level for transmission, according to the current state, $s_t \in \mathcal{S}$, based on the decision policy $\pi$. The transmission power is discretized into 3 levels, which leads to a $3 \times N_{RB}$ as the dimension of the action space if there are $N_{RB}$ resource blocks.

The reward function relates to two parts: the V2I capacity and the latency constraints. In our settings, the reward remains positive if the constraints are satisfied; it will be a penalty, a negative reward, $r_N < 0$ if the V2V constraints are violated. If the constraints are satisfied at current time slot $t$, then the V2V pair received a positive reward, proportion to the sum of the V2I capacity. Therefore, the reward function can be expressed as,
\begin{equation}
r_t = \begin{cases}
\sum{C_m},& \ \text{if latency constraints are satisfied,}  \\
r_N,& \ \text{otherwise.}
\end{cases}
\end{equation}

The state transition and reward are stochastic and follow the Markov decision process (MDP), where the state transition probabilities and rewards depend only on the state of the environment and the action taken by the agent.
The transition from $s_t$ to $s_{t+1}$ with reward $r_t$ when action $a_t$ is taken can be characterized by the conditional transition probability $p(s_{t+1},r_t|s_t,a_t)$.
Note that the agent can only control its own actions and has no a priori knowledge of transition probability matrix $P = (p(s_{t+1},r_t|s_t,a_t))$, which is determined by the environment.
In our case, the transition on the channels, the interference, and the remaining messages to transmit are generated by the simulator of the wireless environment.
The goal of learning is to maximize the gain defined as the expected cumulative discounted rewards,
\begin{equation}
G_t = \mathbb{E}[\sum_{n=0}^{\infty} \beta^n r_{t+n}],
\end{equation}
where $\beta$ is the discount factor.

\subsection{Q-Learning}
The agent takes actions based on a policy, $\pi$, which is a mapping from the state space, $\mathcal{S}$, to the action space, $\mathcal{A}$, and can be expressed as $\pi: s_t \in \mathcal{S} \rightarrow a_t \in \mathcal{A}$.
As indicated before, the action, $a_t \in \mathcal{A}$, corresponds to how to select power and spectrum given a state $s_t$ described above in our problem.

We use Q-learning to get an optimal policy for resource allocation in V2V communications to maximize the long-term expected accumulated discounted rewards \cite{Deep}.
The Q-value for a given state-action pair $(s_t,a_t)$, $Q(s_t, a_t)$, of policy $\pi$ is defined as the expected accumulated discounted rewards when taking an action $a_t \in \mathcal{A}$ and following policy $\pi$ thereafter.
Once Q-values, $Q(s_t, a_t)$, are given, a policy, $\pi$, can be easily constructed,
\begin{equation}
a_t = \arg\max_{a_t \in \mathcal{A}} Q(s_t,a_t).
\end{equation}
That is, the action is taken with the maximum long-term accumulated rewards.

The optimal policy with Q-values $Q^*$ can be found without any knowledge of the system dynamics based on the following update equation,
\begin{equation}
\begin{aligned}
Q_{new}(s_t, a_t)  = & Q_{old}(s_t, a_t) + \alpha [r_{t+1} +\\
                     &\beta \max_{s\in \mathcal{S}} Q_{old}(s, a_t) - Q_{old}(s
_t, a_t)],
\end{aligned}
\end{equation}
It has been shown that in the MDP case, the Q-values will converge with probability 1 to the optimal $Q^*$ if each action is executed in each state an infinite number of times on an infinite run and the learning rate $\alpha$ decays appropriately.

\subsection{Deep Q Networks}
\begin{figure}[!t]
\centering
\includegraphics[width = 0.9 \linewidth]{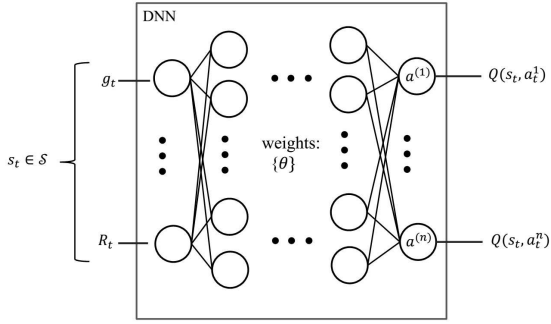}
\caption{Structure of Deep Q-networks}
\end{figure}
Q-learning works well if the state and action spaces of the problem are small and a look-up table can be used to accomplish the update rule.
However, this becomes impossible when the state-action space becomes very large.
In this situation, many states may be rarely visited, thus the corresponding Q-values are seldom updated, leading to a much longer time to converge.
Deep Q-network combines Q-learning with deep learning. The Q-function is approximated by a deep neural network as shown in Fig. 3.
The basic idea behind deep Q-network is the use of a deep neural network (DNN) function approximator with weights \{$\theta$\} as a Q-network \cite{Deep}.
Once \{$\theta$\} is given, Q-values, $Q(s_t, a_t)$ will be determined.
Deep neural networks will address sophisticate mappings between the channels information and the desired output based on a large amount of training data, which will be used to determine Q-values.

The Q-network updates its weights, $\theta$, at each iteration to minimize the following loss function derived from the same Q-network with old weights on a data set $D$,
\begin{equation}
Loss(\theta) = \sum_{(s_t,a_t) \in D}(y - Q(s_t,a_t,\theta))^2,
\end{equation}
where
\begin{equation}
y = r_t+\max_{a \in \mathcal{A}} Q_{old}(s_t,a,\theta),
\end{equation}
where $r_t$ is the corresponding reward.

\subsection{Training and Testing Algorithms}
As most machine learning algorithms, ours consists of two stages in our system, the training stage and the testing stage.
The training and test data are generated from an environment simulator and the agents. Each sample includes $s_t$, $s_{t+1}$, $a_t$, and $r_t$. 
Our simulator consists of DUEs and CUEs and their channels, where the vehicles are randomly dropped and the channels for CUEs and DUEs are generated based on the positions of the vehicles.
With the selected spectrum and power of V2V links, the simulator can provide the $s_{t+1}$ and $r_t$ to the agents.
In the training stage, we follow the deep Q-learning with experience replay \cite{Deep}, where the generated data are saved in a storage called \emph{memory}.
As shown in Algorithm 1, the mini-batch data used for updating the Q-network is sampled from the \emph{memory} in each iteration.
In this way, the temporal correlation of data can be suppressed.
The policy used in each V2V link for selecting spectrum and power is random at the beginning and is gradually improved with the updated Q-networks.
As shown in Algorithm 2, in the test stage, the actions in V2V links are chosen with the maximum Q-value given by the trained Q-networks, based on which the evaluation is obtained.

\begin{algorithm}
\caption{Training Stage Procedure}\label{euclid}
\begin{algorithmic}[1]
\Procedure{Training}{}\\
\textbf{Input}: Q-network structure, environment simulator.\\
\textbf{Output}: Q-network\\

\textbf{Start:}

Random initialize the policy $\pi$

Initialize the model

Start environment simulator, generate vehicles, V2V links, V2I links.\\
\textbf{Loop}:

Random sample V2V links in the system.

Generate a set of data using policy $\pi$ from the environment simulator.

Save the data item \{state, reward, action, post-state\} into memory.

Sample a mini-batch of data from the memory.

Train the deep Q-network using the mini-batch data.

Update the policy: chose the action with maximum Q-value.\\

\textbf{End Loop}\\

\textbf{Return}: Return the deep Q-network

\EndProcedure
\end{algorithmic}
\end{algorithm}

\begin{algorithm}
\caption{Test Stage Procedure}
\begin{algorithmic}[1]
\Procedure{Testing}{}\\
\textbf{Input}: Q-network, environment simulator.\\
\textbf{Output}: Evaluation results\\
\textbf{Start:}
Load the Q-network model

Start environment simulator, generate vehicles, V2V links, V2I links.\\

\textbf{Loop}:

Random sample V2V links in the system.

Select the action by choosing the action with the largest Q-value.

Update the environment simulator based on the actions selected.

Update the evaluation results, i.e., the average of V2I capacity and the probability of successful DUEs.
\\

\textbf{End Loop}\\
\textbf{Return}: Evaluation results

\EndProcedure
\end{algorithmic}
\end{algorithm}

\section{Simulations}

In this section, we present simulation results to demonstrate the performance of the proposed method.
We consider a single cell outdoor system with the carrier frequency of $2$ GHz.
We follow the simulation setup for the Manhattan case detailed in 3GPP TR $36.885$ \cite{3GPP}, where there are 9 blocks in all and with both line-of-sight (LOS) and non-line-of-sight (NLOS) channels.
The vehicles are dropped in the lane randomly according to the spatial Poisson process and each plans to communicate with the nearby $3$ vehicles.
Our deep Q-network is a five-layer fully connected neural network with three hidden layers.
The numbers of neurons in the three hidden layers are 500, 250 and 120, respectively.
The activation function of Relu is used, which is defined as
\begin{equation}
f_r(x) = \max(0,x).
\end{equation}
We also utilize $\epsilon$-greedy policy to balance the exploration and exploitation \cite{Deep} and adaptive moment estimation method (Adam) for training \cite{ADMM}. The detail parameters can be found in Table 1.

\begin{table}
\centering
\caption{Simulation Parameters}
\begin{tabular}{|c|c|}
\hline
\rowcolor{gray}
Parameter & Value\\    \hline
Carrier frequency & 2 GHz \\ \hline
Bandwidth & 10 MHz \\ \hline
BS antenna height & 25m \\ \hline
BS antenna gain & 8dBi\\ \hline
Vehicle speed & 36 km/h\\ \hline
Number of lanes & 3 in each direction (12 in total)\\ \hline
Latency constraints for V2V links & 100 ms\\ \hline
V2V transmit power& 23 dBm\\ \hline
Noise power $\sigma^2$ & -114 dBm\\ \hline
Penalty of latency constraint $P$ & -20 \\ \hline

\end{tabular}
\label{tab:default}

\end{table}

The proposed method is compared with other two methods.
The first is a random resource allocation method.
At each time, the agent randomly chooses a sub-band for transmission.
The other method is developed in \cite{baseline}, where vehicles are first grouped by the similarities and then the sub-bands are allocated and adjusted iteratively in each group.

\begin{figure}[!t]
\centering
\includegraphics[width=1\linewidth]{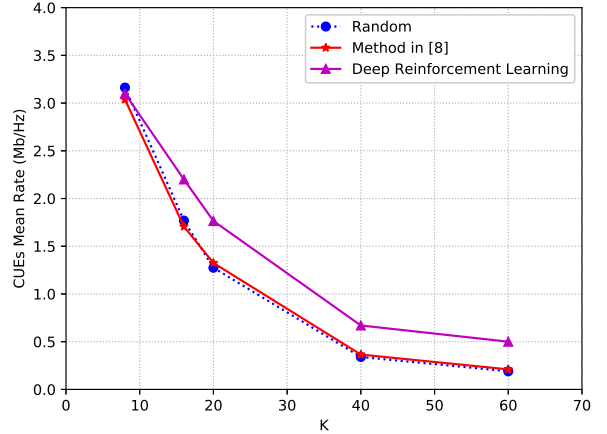}
\caption{Mean rate versus the number of V2V links. }
\end{figure}

\subsection{V2I Capacity}
Fig. 4 shows the average V2I rate versus the number of V2V links.
From the figure, the proposed method has a much better performance to mitigate the interference of V2V links to the V2I communications.
Since the method in \cite{baseline} maximizes the SINR in V2V links, rather than optimizing the V2I links directly, leading to only a slightly better performance than the random method, much worse than the proposed method.

\begin{figure}[!t]
\centering
\includegraphics[width=1\linewidth]{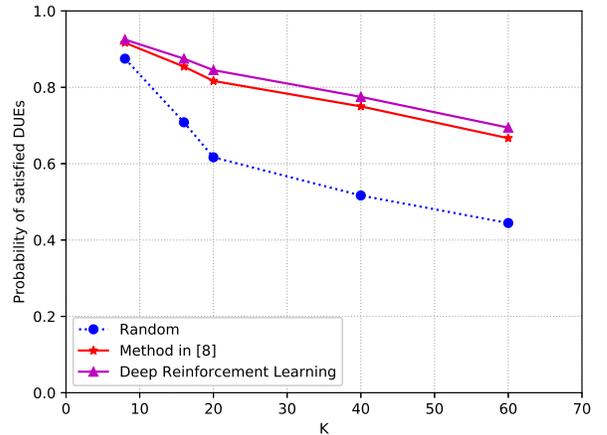}
\caption{Probability of successful versus the number of V2V links.}
\end{figure}

\subsection{V2V Latency}
Fig. 5 shows the probability that V2V links satisfy the latency constraint versus the number of V2V links, $K$. From the figure, the proposed method has a larger probability for DUEs to satisfy the latency constraint since it can dynamically adjust the power and sub-band for transmission so that the links likely violating the latency constraint have more resources.



\section{Conclusion}

In this article, we develop a decentralized resource allocation mechanism for the V2V communication systems based on deep reinforcement learning.
Each V2V link is regarded as an agent, making its own decisions to find optimal sub-band and power level for transmission.
Since the proposed method is decentralized, the global information is not required for each agent to make its decisions, the transmission overhead is small.
From the simulation results, each agent can learn how to satisfy the V2V constraint while minimizing the interference to V2I communications.

\section{Acknowledgment}
The authors would like to thank Dr. May Wu, Dr. Satish C. Jha, Dr. Kathiravetpillai Sivanesan, and Dr. JoonBeom Kim from Intel Corporation for their insightful comments, which have substantially improved the quality of this paper.

\bibliographystyle{IEEEbib}
\bibliography{strings,refs}

\end{document}